\documentclass[prl, superscriptaddress ,twocolumn]{revtex4}
\usepackage{amsmath}
\usepackage{amsfonts}
\usepackage{graphicx}
\usepackage{hyperref}
\usepackage{braket}
\usepackage{gensymb}

\begin{document}

\title{Selective probing of hidden spin-polarized states in inversion-symmetric bulk MoS$_2$}

\author{E. Razzoli}
\email[Corresponding Author: ]{elia.razzoli@unifr.ch}
\altaffiliation[\\Present address: ]
{Quantum Matter Institute, Department of Physics and Astronomy, University of British Columbia, Vancouver, British Columbia V6T 1Z1, Canada}
\author{T. Jaouen}
\author{M.-L. Mottas}
\author{B. Hildebrand}
\author{G. Monney}
\affiliation{D{\'e}partement de Physique and Fribourg Center for Nanomaterials, Universit\'e de Fribourg, CH-1700 Fribourg, Switzerland}
\author{A. Pisoni}
\affiliation{Laboratory of Physics of Complex Matter, Ecole Polytechnique F{\'e}d{\'e}rale de Lausanne, CH-1015 Lausanne, Switzerland}
\author{S. Muff}
\author{M. Fanciulli}
\affiliation{Swiss Light Source, Paul Scherrer Institute, CH-5232 Villigen PSI, Switzerland}
\affiliation{Institute of Physics, {\'E}cole Polytechnique F{\'e}d{\'e}rale de Lausanne, CH-1015 Lausanne, Switzerland}
\author{N. C. Plumb}
\author{V. A. Rogalev}
\author{V. N. Strocov}
\affiliation{Swiss Light Source, Paul Scherrer Institute, CH-5232 Villigen PSI, Switzerland}
\author{J. Mesot}
\affiliation{Swiss Light Source, Paul Scherrer Institute, CH-5232 Villigen PSI, Switzerland}
\affiliation{Institute of Physics, {\'E}cole Polytechnique F{\'e}d{\'e}rale de Lausanne, CH-1015 Lausanne, Switzerland}
\author{M. Shi}
\affiliation{Swiss Light Source, Paul Scherrer Institute, CH-5232 Villigen PSI, Switzerland}
\author{J. H. Dil}
\affiliation{Swiss Light Source, Paul Scherrer Institute, CH-5232 Villigen PSI, Switzerland}
\affiliation{Institute of Physics, {\'E}cole Polytechnique F{\'e}d{\'e}rale de Lausanne, CH-1015 Lausanne, Switzerland}
\author{H. Beck}
\author{P. Aebi}
\affiliation{D{\'e}partement de Physique and Fribourg Center for Nanomaterials, Universit\'e de Fribourg, CH-1700 Fribourg, Switzerland}

\date{\today}

\begin{abstract}
Spin- and angle-resolved photoemission spectroscopy is used to reveal that a large spin polarization is observable in the bulk centrosymmetric transition metal dichalcogenide MoS$_2$. It is found that the measured spin polarization can be reversed by changing the handedness of incident circularly-polarized light. Calculations based on a three-step model of photoemission show that the valley and layer-locked spin-polarized electronic states can be selectively addressed by circularly-polarized light, therefore providing a novel route to probe these hidden spin-polarized states in inversion-symmetric systems as predicted by Zhang \textit{et al.} [Nature Physics \textbf{10}, 387 (2014)].
\end{abstract}

\maketitle

Transition metal dichalcogenide (TMDC) monolayers have been heavily investigated due to the locking of the spin with valley pseudospins and the presence of a direct gap, which makes them ideal candidates for valleytronic devices \cite{ Mak2016}. Thanks to the lack of inversion symmetry and the non-negligible spin-orbit coupling, TMDC monolayers also feature well-defined spin-polarized ground states \cite{Xiao2012}, which can in principle be investigated by spin- and angle-resolved photoemission spectroscopy (spin-ARPES). However, while few ARPES studies clearly observed the indirect to direct band gap transition going from the bulk crystal to the monolayer \cite{Jin2013, Zhang2014NatNano, Eknapakul2014, Alidoust2014}, spin-ARPES investigation of TMDC monolayers is more challenging, given the low cross-section of photoemission from single layers. 

ARPES measurements on the bulk system are instead less demanding but early studies detected spin-resolved signals only from TMDCs with broken inversion symmetry \cite{Suzuki2014}. Interestingly, a recent theoretical study \cite{Zhang2014NatPhys} suggested that the spin texture of the TMDC could be probed by photoemission, even in the inversion symmetric bulk TMDC crystals, as a result of the localization of two spin-degenerated valence band maxima on different layers of the unit cell and of the finite penetration depth of the photoemission process probing preferentially the uppermost layer. Experimentally this effect has been observed for WSe$_2$ where the spin orbit (SO) coupling is the strongest one among the TMDCs \cite{Riley2014}.
   
Here we show that a large out-of-plane spin-polarization is observable in the bulk dichalcogenide MoS$_2$, and more importantly, that its sign depends on the handedness of the incident circularly-polarized light. Our calculations, based on a three step model of the photoemission process demonstrate that the observed spin reversal is an initial state effect. Using left ($C_L$) and right-handed ($C_R$) circularly-polarized light results in selecting different initial states that present a positive or negative out-of-plane spin polarization depending on their localization on the S-Mo-S layers of the 2$H$-stacked MoS$_2$ unit cell. Our findings not only highlight the locking of the spin with layer and valley pseudospins in MoS$_2$ but also provide a novel and improved route, other than taking profit of the inelastic mean free path (IMFP) \cite{Zhang2014NatPhys}, for selectively probing hidden spin-polarized bands in inversion symmetric systems. 

\begin{figure*}[t]
\includegraphics[width=1\textwidth]{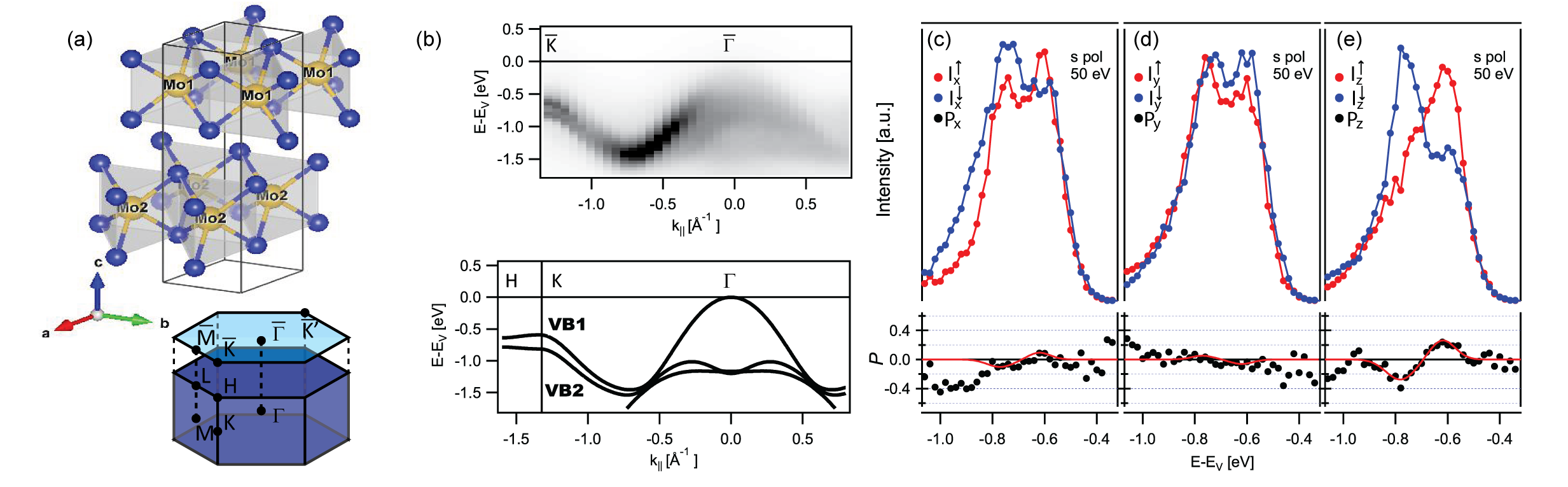}
\caption{ (a) Structural model (top) and Brillouin zone (bottom) of 2$H$-MoS$_2$. (b) ARPES intensity map along $\overline{K}-\overline{\Gamma}$ direction (top), and DFT-LDA calculated band structure showing the band dispersion along $K-\Gamma$ and $H-K$ directions (bottom). (c)-(d) Spin-resolved intensities (top) and spin polarization curves (bottom) at the  $\overline{K}$-point along $x$, $y$, and $z$, respectively. Continuous red lines in the bottom panels are Gaussian fits, resulting in $P_x= \mp 0.08 $, $P_y = \pm 0.05$ and $P_z = \mp 0.28$. Spin-ARPES data were acquired at $h\nu= 50$ eV with $s-$polarized light.
}\label{InPlane}
\end{figure*}

The soft x-ray ARPES and spin-ARPES experiments were respectively performed at the ADRESS beamline \cite{Strocov2014} and at the COPHEE end-station of SIS beamline at the Swiss Light Source, Paul Scherrer Institut, Villigen, Switzerland. Base pressures were better than $4 \times 10^{-10}$ mbar, with a sample temperature of $\sim$25 K \cite{Supplementary}. The chemical vapor transport-grown 2$H$-MoS$_2$ crystal consisting of hexagonal 2$H$-stacked S-Mo-S layers separated by a van der Waals gap (Fig. 1(a) \cite{Vesta}), has been cleaved \textit{in-situ} at low temperature. Figure 1(b) (top panel) shows the ARPES intensity plot of bulk MoS$_2$ along the $\overline{K}$-$\overline{\Gamma}$ high-symmetry direction of the Brillouin zone [bottom of Fig. 1(a)] and measured using circularly-polarized light with an overall energy resolution of 50-80 meV. The observed bands are in agreement with previous ARPES measurements \cite{Jin2013, Eknapakul2014, Latzke}, with the top of the valence band at $\overline{\Gamma}$ originating from Mo $d_{z^2}/p_z$ orbitals and that lies $\sim$600 and $\sim$800 meV above the upper (VB1) and lower (VB2) valence bands at $\overline{K}$ with mainly Mo $d_{x^2-y^2}$, $d_{xy}$ orbital characters. The observed dispersion is in good agreement with our density functional theory (DFT) calculation within the local density approximation (LDA) between $\Gamma$ and $K$, given the small $k_z-$dispersion along $K-H$ direction (see Fig. 1(b) bottom panel) \cite{Supplementary}.

Figures 1 (c)-(e) present the experimental spin-ARPES intensities (top) and the spin-polarization energy distribution curves (bottom) at the $\overline{K}$ point along the three spatial coordinates $x$, $y$, and $z$ in the sample frame measuring in-plane ($P_x$, $P_y$), and out-of-plane ($P_z$) spin polarization. Conversion from the coordinate system given by the
Mott polarimeters to the sample frame is performed according to \cite{Meier2008}.  The difference in binding energy between VB1 and VB2 is 170 meV, in agreement with previous studies \cite{Latzke}. 
The measured spin polarization is almost entirely out-of-plane with a negligible small in-plane component. The measured out-of-plane spin polarization points upwards and downwards for the upper and lower valence band, respectively, similarly to what has been observed in WSe$_2$ \cite{Riley2014}.

\begin{figure*}[t]
\includegraphics[width=1
\textwidth]{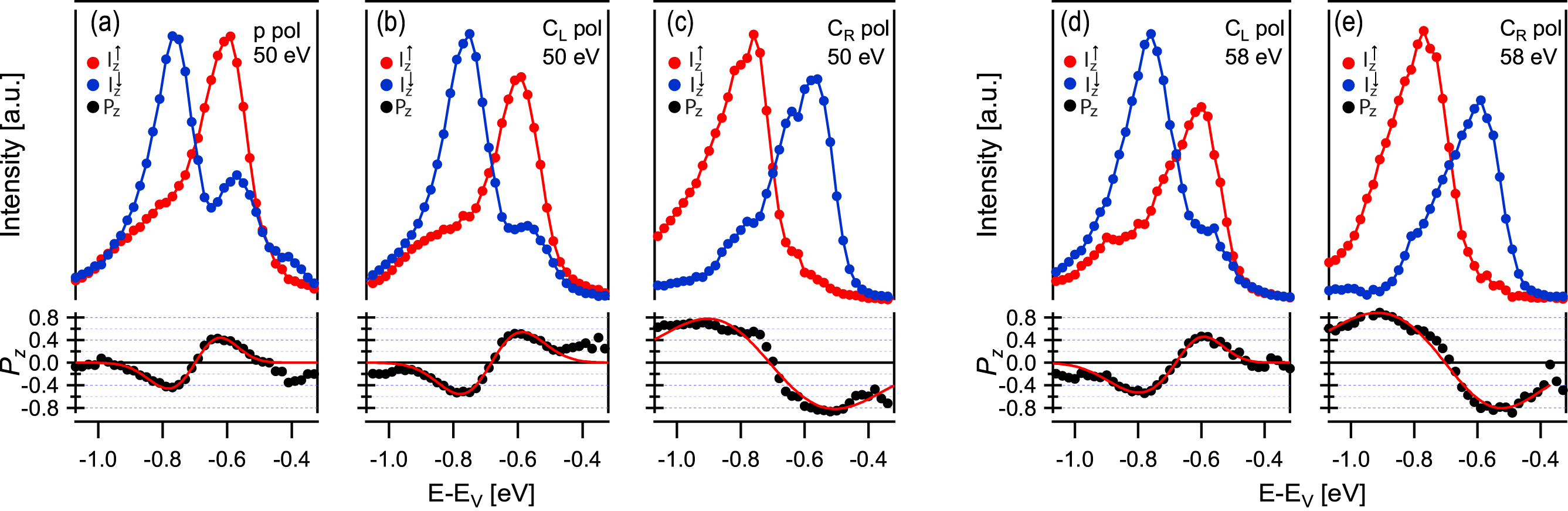}
\caption{ Spin-ARPES data acquired at $h\nu= 50 $ and $h\nu= 58 $ eV. 
(a)-(c) Spin-resolved intensities (top) and spin polarization curves (bottom) along $z$ at the $\overline{K}$-point acquired with p, $C_L$ and $C_R$ polarization, respectively ($h\nu= 50 $ eV). (d), (e) Spin-resolved intensities (top) and spin polarization curves (bottom) at the $\overline{K}$-point acquired with $C_L$ and $C_R$ polarization, and $h\nu= 58 $ eV. Continuous red lines in the bottom panels are Gaussian fits, resulting in $P_z(p, 50 \text{ eV})= \mp 0.50 $, $P_z(C_L, 50  \text{ eV}) = \mp 0.68$, $P_z(C_R, 50  \text{ eV}) = \pm 0.85$, $P_z(C_L, 58  \text{ eV}) = \mp 0.54$, and $P_z(C_R, 58  \text{ eV}) = \pm 0.91$.
}\label{PzExp}
\end{figure*}

The $z$-component of the spin polarization displays an interesting polarization dependence [Fig. \ref{PzExp}]. The spin polarization measured with $p$- [Fig. \ref{PzExp} (a)], and $C_L$ polarization [Fig. \ref{PzExp} (b)] are in agreement with the observation with $s$-polarized light [Fig. 1 (e)], with the only difference of an increasing absolute value of the spin polarization fraction.
Instead, when probed with $C_R$ light the out-of-plane spin polarization changes its sign, i.e., the $z$-component of the detected photo-electron points upwards and downwards for the lower and upper valence band, respectively [Fig. \ref{PzExp} (c)]. Interestingly, the light handedness-induced reversal of the $z$-component is also observable at other photon-energy [e.g. at $h \nu =58$ eV as shown in Fig. \ref{PzExp} (d)-(e)], therefore indicating that it does not depend on the three-dimensional ($k_z$) band dispersion.

A possible explanation of our observation would be that photoelectron spin flips during the photoemission process, i.e., we observed a final state matrix element effect. Up to now, most of spin-ARPES studies dealing with $k$-resolved light-induced spin rotations have been carried out on topological insulators (TI), due to the large strength of the SO-coupling and to the high quality of the ARPES data in such systems \cite{Zhu2013, Zhu2014, Jozwiak2013, Barriga2014}. It has been experimentally shown that light-induced spin rotations are observable only at low photon energy (about 6 eV) \cite{Jozwiak2013, Barriga2014}. Theoretically it has been suggested that the SO-induced term in the light-matter interaction Hamiltonian $H^{int}_{SO}$ \cite{ParkLouie2012} is responsible for the spin events tuned by light polarization. However, in contrast to the TI that exhibit extremely large SO-coupling, in TMDCs and especially in MoS$_2$ \cite{Eknapakul2014}, the $\braket{i|H^{int}_{SO}|f}$ term can be neglected and the observed flip in the out-of-plane spin-polarization has to be related to a different mechanism \cite{Supplementary}.

Instead, the origin of the sign inversion of $P_z$ can be understood analyzing the idealized case of a TMDC bilayer with small interlayer coupling $t_{\perp}$. In this simplified model, the top of the valence band VB1 ($E_b=E_1$) at $K$ is doubly degenerate and originates in Mo $d_{x^2-y^2}$, $d_{xy}$ orbitals. The eigenstates expressed in terms of spherical harmonics $\ket{l,m}_{\mu}$ are then given by \cite{Supplementary, Gong2013}:
\begin{align}
\label{InitialStateBi}
\begin{split}
\ket{1, E_1} &= \frac{\cos\alpha  \ket{2,2}_{1} + \sin\alpha \ket{2,-2}_{2}     }{\sqrt{2}}  \ket{\uparrow}   \\
\ket{2, E_1} &= \frac{ \sin \alpha \ket{2,2}_{1} + \cos \alpha  \ket{2,-2}_{2}}{\sqrt{2}} \ket{\downarrow}
\end{split}
\end{align}
using $\mu=1(2)$ for the upper (lower) Mo layer of the 2$H$-stacked MoS$_2$ unit cell [Fig. 3(a)], and with $\cos 2\alpha = \lambda_{so}/\sqrt{\lambda^2_{so}+t^2_{\perp}}$ that corresponds to the spin-dependent layer polarization of the holes states. The spin-valley coupling of holes in monolayers $\lambda_{so}$ is calculated to be 1.7 times higher than the interlayer hopping $t_{\perp}$ in MoS$_2$ \cite{Gong2013}, therefore resulting in a substantial layer-dependent spin-polarization ($\cos 2\alpha = 0.863$).

The states $\ket{1, E_1}$ and $\ket{2, E_1}$ respectively contribute to the spin-up ($I^{\boldsymbol\epsilon}_{\uparrow,z}$) and spin-down ($I^{\boldsymbol\epsilon}_{\downarrow,z}$) intensity \cite{Supplementary}:
\begin{align}
\begin{split} \label{Ibi}
I^{\boldsymbol\epsilon}_{\uparrow,z} \propto | \cos(\alpha)  {\boldsymbol\kappa}_{2,2} \cdot \boldsymbol\epsilon  + \sin(\alpha) \delta e^{-i \textbf{k}\cdot \textbf{d}_l } {\boldsymbol\kappa}_{2,-2} \cdot \boldsymbol\epsilon |^2 \\
I^{\boldsymbol\epsilon}_{\downarrow,z} \propto |\sin(\alpha) {\boldsymbol\kappa}_{2,2} \cdot \boldsymbol\epsilon  + \cos(\alpha) \delta  e^{-i \textbf{k}\cdot \textbf{d}_l } {\boldsymbol\kappa}_{2,-2} \cdot \boldsymbol\epsilon|^2 
\end{split}
\end{align}
with $\boldsymbol\kappa_{l,m}=\braket{e^{i\textbf{k} \cdot \textbf{r}}|\textbf{r}|l,m}$, $\textbf{d}_l=(\frac{1}{2}a,\frac{2}{3}a,-\frac{1}{2}c)$ the relative atomic displacement between the upper and lower Mo layers, $\boldsymbol\epsilon$ the light polarization vector, and $\delta=e^{-\frac{c}{2\lambda_e}}$, an attenuation factor that takes into account the photoelectrons IMFP $\lambda_e$. For normal incident circularly-polarized light the conservation of the total angular momentum dictates that only the states with $m_k= m\pm 1$, can give a non-vanishing contribution to $\boldsymbol \kappa_{l,m}$, once the free-electron final state is decomposed in spherical harmonics $\ket{e^{i\textbf{k}\cdot \textbf{r}}}= \sum_{l_k,m_k} d_{l_k,m_k}(\textbf{k}, r) \ket{l_k,m_k}$. 

In Fig. 3 we illustrate the dipole-allowed transitions for the eigenstate of VB1 at the $K$ point, for the case of vanishing interlayer coupling ($\alpha=0$). In this case spin up and spin down states are fully locked on the upper and lower S-Mo-S layer, respectively [Fig. 3(a)]. 

For $C_L$ light [Fig. 3(b), left-hand-side], the available final states have $m_k=m-1$, and are $\{\ket{1,1} , \ket{3,1}\}$ for $m=2$ and $\{ \ket{3,-3} \}$ for $m=-2$. As expected from the higher number of final states available for a dipole transition from a state with $m=2$, our numerical evaluation shows that the cross-section for transition to the states $\{\ket{1,1} , \ket{3,1}\}$ is one order of magnitude larger than to $\{\ket{3,-3}\}$, i.e.,  $|\boldsymbol \kappa_{2,2}\cdot \boldsymbol\epsilon^{\textbf{n}}_{C_L}| /|\boldsymbol \kappa_{2,-2}\cdot \boldsymbol\epsilon^{\textbf{n}}_{C_L}| \approx 10$ \cite{Supplementary}. Similarly and according to the relation $\boldsymbol \kappa_{l,\pm m}\cdot \boldsymbol\epsilon_{C_{R/L}} = \boldsymbol \kappa_{l,\mp m}\cdot \boldsymbol\epsilon_{C_{L/R}}$, the available final states of a dipole transition induced by $C_R$ light  [Fig. 3(b) right-hand-side], have $m_k=m+1$ and $|\boldsymbol \kappa_{2,-2}\cdot \boldsymbol\epsilon^{\textbf{n}}_{C_R}|/|\boldsymbol \kappa_{2,2}\cdot \boldsymbol\epsilon^{\textbf{n}}_{C_R}| \approx 10$. For a non-zero interlayer hopping, eq. (\ref{Ibi}) thus gives:
\begin{align}
P_z^{C^{\textbf{n}}_{R/L}} = \frac{ I^{\boldsymbol\epsilon^{\textbf{n}}_{C_{R/L}}}_{\uparrow,z}-I^{\boldsymbol\epsilon^{\textbf{n}}_{C_{R/L}}}_{\downarrow,z}}{I^{\boldsymbol\epsilon^{\textbf{n}}_{C_{R/L}}}_{\uparrow,z}+I^{\boldsymbol\epsilon^{\textbf{n}}_{C_{R/L}}}_{\downarrow,z}} \approx \mp \cos 2\alpha. 
\end{align}
%
%
\begin{figure}[b]
\includegraphics[width=0.5\textwidth]{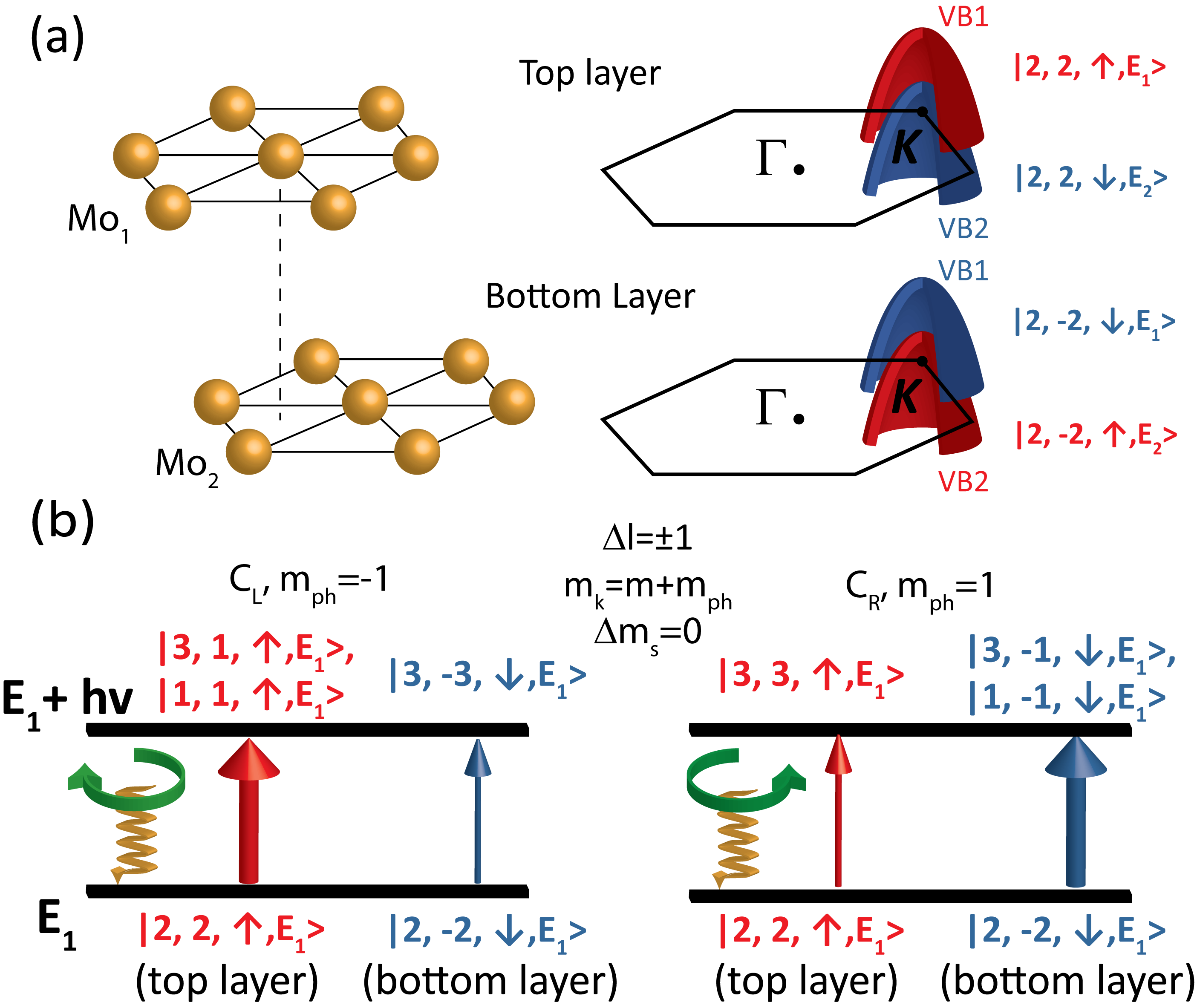}
\caption{  (a) Schematic of the layer-resolved band dispersion of VB1 and VB2 around the $K$ point. (b) Dipole-allowed transitions  for the eigenstate of VB1 for vanishing interlayer coupling ($\alpha=0$) at the $K$ point.
}\label{Sketch}
\end{figure}
%
\begin{figure*}[t]
\includegraphics[width=1\textwidth]{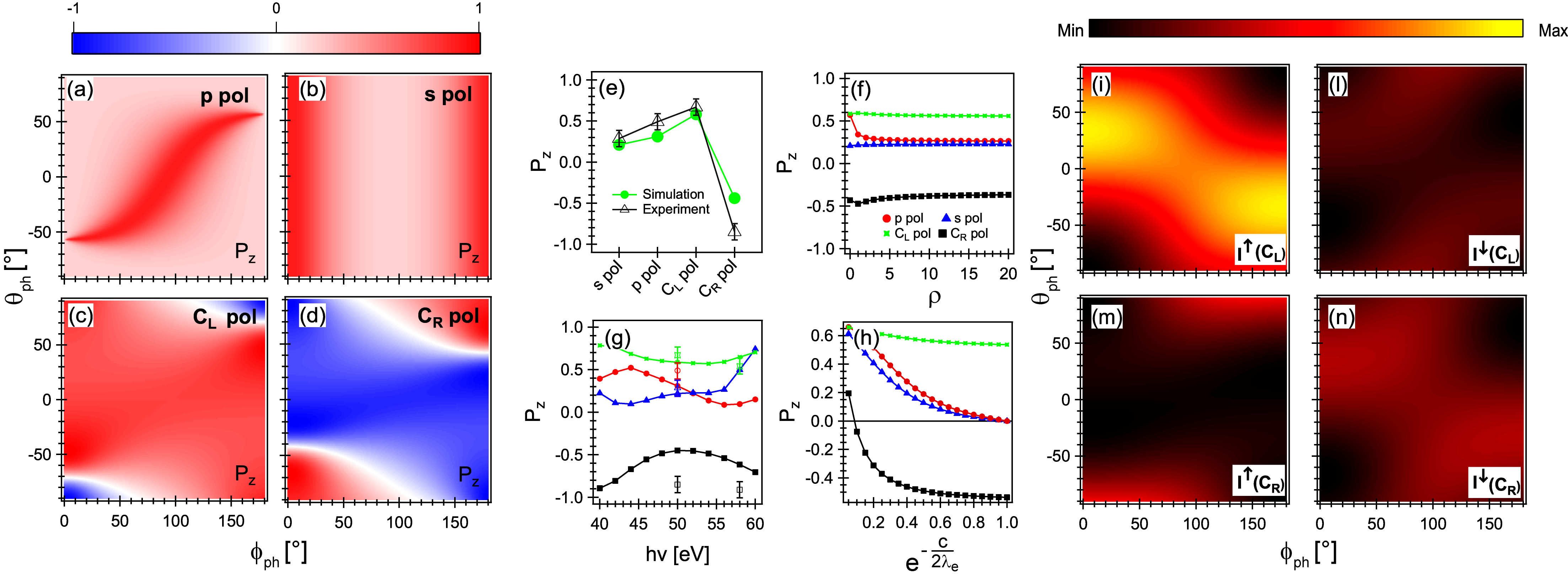}
\caption{Three-step model photoemission calculations. (a)-(d) Spin polarization ($P_z$) maps for  $p$, $s$, $C_L$ and $C_R$ polarization, respectively ($h\nu= 50 $ eV).  (e) Experimental and simulated spin polarization along $z$.  (f)-(h) 	 radial integral ($\rho$), photon energy ($h\nu$) and IMFP ($\lambda_e$)-dependence of $P_z$, respectively. Empty symbols in (g) show the experimental $P_z$ from Fig. 2 (a)-(e). (i)-(n)  Spin-resolved intensity maps for $C_L$ (top panels) and $C_R$ (bottom panels) polarized light.
}\label{Model}
\end{figure*}

This demonstrates that the layer-locked, hidden spin-polarized states of the $K$ valley can be probed through the spin-dependent selection rules introduced by the use of circularly-polarized light. This not only selects which hidden-spin polarized state will be observed, but also overcomes the IMFP-limited probing of the spin texture of the uppermost layer as introduced by using linearly-polarized light \cite{Riley2014, Zhang2014NatPhys, Gehlmann2016, footnote1}.

This interpretation is confirmed by calculating the spin-resolved ARPES intensity and the spin polarization for 8 MoS$_2$ layers as a function of both, the light geometry and polarization and using the spin doublet of the topmost VB (VB1) in valley $K$ as the initial state \cite{Supplementary}. The spin up (down) state is highly localized on the odd (even) Mo layer, in analogy to the bilayer model and $\lambda_e$, included following the methods proposed by \cite{Zhu2013, Zhu2014}, has been fixed to the interlayer distance ($6.1$ \AA ). 

The spin polarization intensity maps as a function of the incoming light geometry and polarization [Fig. \ref{Model} (a)-(d)] \cite{footnote2}, show that for $s-$ and $p$-polarized light the $z$-component of the spin is positive irrespective of the geometry, in agreement with our experimental observations and with what is expected in MoS$_2$ as the result of the finite probing depth of the photoemission process \cite{Zhang2014NatPhys, Riley2014}. Our model further predicts positive and negative spin polarization for $C_L$ and $C_R$-polarized lights for most of the possible experimental geometries, including ours ($\theta_{ph}=48.3  \degree$ and $\phi_{ph}=71  \degree$), as well as sign inversion for extreme geometries corresponding to grazing incidence of the radiation [Fig. \ref{Model} (c), (d)].   

Overall, the theoretical $P_z$ values obtained for the various light polarization and corresponding to our geometry agree well with the experimental data as shown in Fig. \ref{Model} (e). The weak dependence of the spin-polarization on the radial integral of the overlap between the initial and the final state ($\rho$) indicates that the calculations are not strongly dependent on the numerical details of the chosen states [Fig. \ref{Model} (f)], and the observed spin-flip is predicted to be robust against changes in the photon energy [Fig. \ref{Model} (g)], as experimentally observed [Fig. \ref{PzExp} (b)-(e)]. The dependence of the calculated $P_z$ on $\lambda_e$ [Fig. \ref{Model} (h)] also confirms that using circularly-polarized light allows overcoming the IMFP limitation in variance with the case of linearly polarized lights where a presence of a small $\lambda_e$ is essential for observing any spin polarization. Finally, the spin-resolved intensity maps in Fig. \ref{Model} (i)-(n) account for the selective ability of the circularly-polarized light to probe the layer-locked spin-polarized states for most of the possible light geometries. As the result of the finite $\lambda_e$ at $h\nu=50$ eV, they also show higher ARPES intensities for $C_L$ than for $C_R$ [Fig. \ref{Model} (i), (n) and (l), (m)], with a theoretical $I^{\uparrow}(C_L)/I^{\downarrow}(C_R)$ ratio of $2.65$, in close agreement with the experimental value of $2.1$ \cite{footnote3}.

We finally note that putative interference effects between distinct degenerate initial states \cite{Heinzmann2012} 
and/or small correction to eq.$(1)$, resulting from  possible modification within the MoS$_2$ monolayer, such as mirror symmetry breaking and VB1-VB2 intervalence mixing, might contribute to the spin-polarization. In bulk TMDCs, these contributions only slightly change $P_z$ but may substantially affect the in-plane spin-polarization ($P_{\parallel}$) \cite{Supplementary}. Similar to previous studies \cite{Riley2014}, we observe a small  $P_\parallel$ [Fig. 1 (c)-(d)], which becomes more sizeable with other polarization \cite{Supplementary}. For supported TMDC layers the interaction with the substrate may in some cases enhance these effects whose contributions to both $P_z$ and $P_\parallel$ might become more important \cite{Mo2016}.

In conclusion our ARPES and spin-ARPES measurements of bulk centrosymmetric 2$H$-MoS$_2$ have revealed a large out-of-plane spin-polarization with sign depending on the handedness of incident circularly-polarized light. Our calculations, based on a three-step model of the photoemission process demonstrate that this relates to an initial-state effect of photoemission through the dipole selection rules that allow to intimately probe the valley and layer spin texture of the electronic states. Our finding provides a novel route for studying hidden spin-polarized bands of inversion symmetric systems in laser- or soft x-ray-ARPES where the inelastic mean free path of photoelectrons is typically large.

\begin{acknowledgments}
We acknowledge F. Boschini,  M. Michiardi and A. Damascelli for valuable discussions. This project was supported by the Fonds National Suisse pour la Recherche Scientifique through Div. II. E.R. acknowledges support from the Swiss National Science
Foundation (SNSF) grant no. P300P2$\_$164649.
\end{acknowledgments}



\begin{thebibliography}{1}
\bibitem{footnote1} For the case of normal incident linearly-polarized light all states with $m_k-m=0, \pm1$ are equally accessible, this result in $|\boldsymbol \kappa_{2,-2}\cdot \boldsymbol\epsilon_{p/s}| \approx |\boldsymbol \kappa_{2,2}\cdot \boldsymbol\epsilon_{p/s}|$, and a positive spin polarization  is observable only when a disproportion of intensity from the two initial states is introduced by the IMFP.
\bibitem{footnote2} The $P_z$ maps have been calculated according to the formulas $P= (I^{\uparrow}-I^{\downarrow})/(I^{\uparrow}+I^{\downarrow}) ,  \quad
I^{\sigma}=  I^{\sigma}{(1)}+ I^{\sigma}{(2)}$, 
where the total photoemission intensity $I^{\sigma}$ associated with a dipole transition to a final state of spin $\sigma$ contains the contributions of the two degenerate initial states $I^{\sigma}{(1)}$ and $I^{\sigma}{(2)}$.
\bibitem{footnote3} In case of finite $\lambda_e$ the magnitude of $I^\uparrow$ is similar for $s$, $p$, and $C_L$ polarizations since it is mainly determined by the photoemission from the first layer. The magnitude of $I^\downarrow$ for $C_R$ is reduced by a factor $\sim e^{-\frac{c}{2 \lambda_e}}$, since it is mainly determined by the photoemission from the second layer.
\end{thebibliography}
\end{document}